  \documentclass[sigconf]{acmart}

\copyrightyear{2021}
\acmYear{2021}
\setcopyright{iw3c2w3}
\acmConference[WWW '21]{Proceedings of the Web Conference 2021}{April 19--23, 2021}{Ljubljana, Slovenia}
\acmBooktitle{Proceedings of the Web Conference 2021 (WWW '21), April 19--23, 2021, Ljubljana, Slovenia}
\acmPrice{}
\acmDOI{10.1145/3442381.3450077}
\acmISBN{978-1-4503-8312-7/21/04}

\acmSubmissionID{fp5184702}

\usepackage[utf8]{inputenc}
\usepackage{hyperref}
\usepackage{url}
\usepackage{fullpage}
\usepackage{subcaption}

\def\Snospace~{\S{}}

\makeatletter
  \newcommand\EatSpacesHack{\@bsphack\@esphack}
\makeatother
  \renewcommand{\comment}[1]{\EatSpacesHack}
  \newcommand{\PostSubmission}[1]{\EatSpacesHack}
  \newcommand{\todo}[1]{\EatSpacesHack}
  \newcommand{\basi}[1]{\EatSpacesHack}
  \newcommand{\ak}[1]{\EatSpacesHack}
  \newcommand{\reviewfix}[1]{\EatSpacesHack} 

\begin{document}

\title{Auditing for Discrimination in Algorithms Delivering Job Ads}

\author{Basileal Imana}
\affiliation{%
  \institution{University of Southern California}
  \city{Los Angeles}
  \state{CA}
  \postcode{43017-6221}\country{USA}
}
\author{Aleksandra Korolova}
\affiliation{%
  \institution{University of Southern California}
  \city{Los Angeles}
  \state{CA}
  \postcode{43017-6221}\country{USA}
}
\author{John Heidemann}
\affiliation{%
  \institution{USC/Information Science Institute}
  \city{Los Angeles}
  \state{CA}
  \postcode{43017-6221}\country{USA}
}

\begin{abstract}
Ad platforms such as Facebook, Google and LinkedIn promise value for advertisers 
  through their targeted advertising.
However,
  multiple studies have shown that ad delivery on such platforms can be skewed by gender or race
  due to hidden algorithmic optimization by the platforms,
  even when not requested by the advertisers.
Building on prior work measuring skew in ad delivery,
  we develop a new methodology for black-box auditing of algorithms for
  \emph{discrimination} in the delivery of \emph{job advertisements}.
Our first contribution is to identify the distinction between skew in ad delivery 
  due to protected categories such as gender or race,
  from skew due to differences in qualification among people in the targeted audience.
This distinction is important in U.S.~law,
  where ads may be targeted based on qualifications,
  but not on protected categories.
Second, we develop an auditing methodology that distinguishes between
  skew explainable by differences in qualifications
  from other factors,
  such as the ad platform's optimization for engagement or training its algorithms on biased data.
Our method controls for job qualification by comparing ad delivery of two concurrent ads
  for similar jobs,
  but for a pair of companies with different de facto gender distributions of employees.
We describe the careful statistical tests
  that establish evidence of non-qualification skew in the results.
Third, we apply our proposed methodology to two prominent targeted advertising
  platforms for job ads:
  Facebook and LinkedIn.
We confirm skew by gender in ad delivery on Facebook,
  and show that it cannot be justified by differences in qualifications.
We fail to find skew in ad delivery on LinkedIn.
Finally, we suggest improvements to ad platform practices that could make external auditing
   of their algorithms in the public interest more feasible and accurate.
\end{abstract}


\maketitle

\section{Introduction}

Digital platforms and social networks have become popular
  means for advertising to users.
These platforms provide many mechanisms that
  enable advertisers to target a specific audience, i.e. specify the criteria that the member to whom an ad is shown should satisfy.
Based on the advertiser's chosen parameters, the platforms
  employ optimization algorithms to decide who sees which ad
  and the advertiser's payments.

Ad platforms such as Facebook and LinkedIn use an automated algorithm to deliver
  ads to a subset of the targeted audience.
Every time a member visits their site or app, the platforms
  run an ad auction among advertisers who are targeting
  that member.
In addition to the advertiser's chosen parameters, such as a bid or budget,
  the auction takes into account  
  an ad \emph{relevance score}, which is based on the ad's predicted engagement level and value to the user.
For example, from LinkedIn's documentation~\cite{QualityScores}:
  ``scores are calculated ... based on your
  predicted campaign performance and the predicted performance of top
  campaigns competing for the same audience.''
Relevance scores are computed by ad platforms
  using algorithms; both the algorithms and the inputs they consider are proprietary.
We refer to the algorithmic process run by platforms
  to determine who sees which ad 
  as \emph{ad delivery optimization}.

Prior work has hypothesized that ad delivery optimization plays
  a role in skewing recipient distribution by gender or race even when the advertiser targets their ad inclusively~\cite{Speicher2018, Sweeney2013, Datta2015, Lambrecht2016}.
This hypothesis was confirmed, at least for Facebook, in a recent study~\cite{Ali2019a}, which showed that for jobs such as lumberjack and taxi driver, Facebook delivered ads to audiences skewed along gender and racial lines, even when the advertiser was targeting a gender- and race-balanced audience.
The Facebook study~\cite{Ali2019a} established that the skew is not due to advertiser targeting or competition from other advertisers, and hypothesized that it could stem from the proprietary ad delivery algorithms trained on biased data optimizing for the platform's objectives~(\autoref{sec:sources_of_skew}). 

Our work focuses on developing an auditing methodology
  for measuring skew in the delivery of \emph{job ads},
  an area where U.S. law prohibits discrimination based on certain attributes~\cite{TitleVII, EEOC}.
We focus on expanding the prior auditing methodology of~\cite{Ali2019a} to bridge the gap
  between audit studies that demonstrate that a platform's ad delivery algorithm results in skewed delivery
  and studies that provide evidence that the skewed delivery is discriminatory,
  thus bringing the set of audit studies one step closer to potential use by regulators to enforce the law in practice~\cite{datta2018discrimination}. 
We identify one such gap in the context of job advertisements: controlling for bona fide occupational qualifications~\cite{TitleVII}
  and develop a methodology to address it.
We focus on designing a methodology that assumes no special access beyond what a regular advertiser sees,
  because we believe that auditing of ad platforms in the public interest needs to be possible by third-parties --- and society should not depend solely on the limited capabilities of federal commissions or self-policing by the platforms.

Our first contribution is to examine how the occupational qualification
  of an ad's audience affects the legal liability an ad platform might incur
  with respect to discriminatory advertising (\autoref{sec:problem_statement}).
Building upon legal analysis in prior work~\cite{datta2018discrimination},
  we make an additional distinction between skew that is due to
  a difference in occupational qualifications among the members of the targeted ad audience,
  and skew that is due to (implicit or explicit use of) protected categories such as gender or race by the platform's algorithms.
This distinction is relevant because U.S. law allows differential delivery
  that is justified by differences in qualifications~\cite{TitleVII},
 an argument that platforms are likely to use to defend themselves
  against legal liability when presented with evidence from audit studies
  such as~\cite{Ali2019a, Speicher2018, Sweeney2013, Datta2015, Lambrecht2016}.

Our second contribution is to propose a novel auditing methodology (\autoref{sec:methodology}) that distinguishes
  between a delivery skew that could be a result of the ad delivery algorithm merely incorporating
  job qualifications of the members of the targeted ad audience from skew due to other
  algorithmic choices that correlate with gender- or racial- factors, but are not related to qualifications.
Like the prior study of Facebook~\cite{Ali2019a},
 to isolate the role of the platform's algorithms
  we control for factors extraneous to the platform's ad delivery choices,
  such as the demographics of people on-line during an ad campaign's run,
  advertisers' targeting, and competition from other advertisers.
Unlike prior work, our methodology relies on simultaneously running \emph{paired} ads for several jobs
  that have \emph{similar qualification requirements}
  but have \emph{skewed de facto (gender) distribution}.
By ``skewed de facto distribution'', we refer to existing societal circumstances that are reflected in
  the skewed (gender) distribution of employees.
An example of such a pair of ads is a delivery driver job at Domino's (a pizza chain) and at Instacart (a grocery delivery service).
Both jobs have similar qualification requirements but one is de facto skewed male
  (pizza delivery) and the other -- female (grocery delivery)~\cite{DominosGender, InstacartGender}.
Comparing the delivery of ads for such pairs of jobs
  ensures skew we may observe can not be attributed to
  differences in qualification among the underlying audience.

Our third contribution is to show that our proposed methodology
  distinguishes between the behavior of ad delivery algorithms
  of different real-world ad platforms,
  and identify those whose delivery skew may be going beyond what is justifiable
  on the basis of qualifications, and thus may be discriminatory (\autoref{sec:experiments}).
We demonstrate this by registering as advertisers and running job ads for real employment opportunities on two platforms,
  Facebook and LinkedIn. 
We apply the same auditing methodology to both platforms and observe contrasting results
  that show statistically significant gender-skew in the case of Facebook,
  but not LinkedIn.

We conclude by providing recommendations for changes that could make auditing of ad platforms more accessible, efficient and accurate
  for public interest researchers (\autoref{sec:recommendations}).

\section{Problem Statement}
  \label{sec:problem_statement}

Our goal is to develop a novel methodology that measures skew in ad delivery
  that is not justifiable on the basis of differences in job qualification requirements in the targeted audience.
Before we focus on qualification,
  we first enumerate the different potential sources of skew
  that need to be taken into consideration when measuring the role of the ad delivery algorithms.
We then discuss how U.S. law may treat qualification as a
  legitimate cause for skewed ad delivery.

We refer to algorithmic decisions by ad platforms that
  result in members of one group being over- or under-represented among the ad recipients as ``skew in ad delivery''.
We consider groups that have been identified as legally protected
  (such as gender, age, race).
We set the baseline population for measuring skew as
  the qualified and available ad platform members targeted by the campaign 
  (see~\autoref{sec:skew} for a quantitative definition).

\subsection{Potential Sources of Skew}
  \label{sec:sources_of_skew}

Our main challenge is to isolate the role of the platform's algorithms in creating skew from 
other factors that affect ad delivery and may be used to explain away any observed skew.
This is a challenge for a third-party auditor because they investigate the platform's algorithms as a black-box,
without access to the code or inputs of the algorithm, or access to the data or behavior of platform members or advertisers.
We assume that the auditor has access only to ad statistics provided by the platform.

Targeted advertising consists of two high-level steps.
The advertiser  \emph{creates} 
  an ad, specifies its target audience, campaign budget, and
  the advertiser's objective.
The platform then \emph{delivers}
  the ad to its users 
  after running an auction among advertisers targeting those users.
We identify four categories of factors that may introduce skew into this process:

First, an advertiser can select \textbf{targeting parameters and an audience}
  that induce skew.
Prior work~\cite{ProPublicaAge, ProPublicaGender, ProPublicaRace, Speicher2018, venkatadri-2020-composition} has shown
  that platforms expose targeting options
  that advertisers can use to create
  discriminatory ad targeting.
Recent changes in platforms have tried to disable such options~\cite{GoogleChanges2020, FbChanges2020a, FbChanges2020b}.

Second, an ad platform can make \textbf{choices in its ad delivery optimization algorithm}
  to maximize ad relevance, engagement, advertiser satisfaction, revenue, or other business objectives, which can implicitly or explicitly result in a skew.
As one example,
  if an image used in an ad receives better engagement from a certain demographic,
  the platform's algorithm may learn this association
  and preferentially show the ad with that image to the subset of the targeted audience belonging to that demographic~\cite{Ali2019a}.
As another example, for a job ad,
  the algorithm may aim to show the ad to users
  whose professional backgrounds better match
  the job ad's qualification requirements.
If the targeted population of qualified individuals is skewed along demographic characteristics,
  the platform's algorithm may propagate this skew in its delivery.

Third, an advertiser's  \textbf{choice of objective}
  can cause a skew.
Ad platforms such as LinkedIn and Facebook support
  advertiser objectives such as reach and conversion.
\emph{Reach} indicates the advertiser wants their
  ad to be shown to as many people as possible in their target audience,
  while for \emph{conversion} the advertiser 
  wants as many ad recipients as possible to take some action, such as clicking through to their site~\cite{LIAdObjectives, FBAdObjectives}.
Different demographic groups
  may have different propensities to take specific actions,
  so a \emph{conversion} objective can implicitly cause skewed delivery.
When the platform's implementation of the advertiser's objective results in a discriminatory skew, the responsibility for it can be a matter of dispute (see \autoref{sec:liability}).  

Finally, there may be 
  \textbf{other confounding factors} that are not under direct control of a particular advertiser or the
  platform leading to skew, such as differing sign-on rates across demographics,
  time-of-day effects, 
  and differing rates of advertiser competition for users from different demographics.
For example, delivery of an ad may be skewed towards men because more
  men were online during the run of the ad campaign, or because competing
  advertisers were bidding higher to reach the women in the audience than to reach the men~\cite{Ali2019a, Lambrecht2016, DI2019}.

In our work, we focus on isolating skew that results
  from an ad delivery algorithm's optimization (the second factor).
Since we are studying job ads, we are interested in further distinguishing skew due to
  an algorithm that incorporates qualification in its optimization from skew that is due to
  an algorithm that perpetuates societal biases without a justification grounded in qualifications.
We are also interested in how job ad delivery is affected by the 
  objective chosen by the advertiser (the third factor).
We discuss our methodology for achieving these goals in~\autoref{sec:methodology}.

\subsection{Discriminatory Job Ads and Liability}
  \label{sec:liability}

Building on a legal analysis in prior work~\cite{datta2018discrimination},
  we next discuss how U.S. anti-discrimination law may treat
  job qualification requirements, optimization objectives,
  and other factors that can cause skew,
  and discuss how the applicability of the law informs the design of our methodology\footnote{We have updated \autoref{sec:liability} after the original submission to WWW '21 to reflect post-camera-ready improvements to our understanding of the legal issues.}.

Our work is unique in underscoring the implications of
  qualification when evaluating potential legal liability ad platforms
  may incur due to skewed job ad delivery.
We also draw attention to the nuances in analyzing the implications
  of the optimization objective an advertiser chooses.
We focus on Title VII, a U.S. law which prohibits preferential or discriminatory
  employment advertising practices using attributes such as gender or race~\cite{TitleVII}.
\basi{}
We interpret this law to apply not just to actions of advertisers but
  also to \emph{outcomes} of ad delivery. 

\comment{}
\basi{}
Title VII allows entities who advertise job opportunities
  to legally show preference based on
  \emph{bona fide occupational qualifications}~\cite{TitleVII},
  which are requirements necessary to carry out a job function.
\basi{}
While it is unclear whether the scope of Title VII applies to ad platforms (as discussed by Datta \emph{et al.}~\cite{datta2018discrimination}),
  to the extent that it may apply,
  it is conceivable that a platform such as Facebook can use qualification as an exception
  to argue that the skew arising from its ad delivery optimization
  does not violate the law.
They may argue that
  skew (shown in prior work~\cite{Ali2019a})
  simply reflects job qualifications.
Therefore,
  our goal is to design an auditing methodology
  that can distinguish between
  skew due to ad platform's use of qualifications
  from skew due to other algorithmic choices by the platform.
\comment{this sen was there:
``To the extent that a platform is liable under Title VII,
  this distinction can eliminate the possibility of platforms using qualification
  as a legal argument against being held liable for discriminatory outcomes when this argument does not apply.''
  but I can't parse it and I think it just (1) repeats what we just said,
  (2) makes a legal argument.
  I'm not sure we're the best to make legal evaluations like this and I suggest we drop this sen.}
\basi{}
The methodology to make such a distinction is one of our main contributions relative
  to prior work.
It also brings findings from audit studies such as ours a step closer to
  having the potential to be used by regulators to enforce the law in practice. 

As discussed in~\autoref{sec:sources_of_skew},
  the objective an advertiser chooses can also be a source of skew.
If different demographic groups tend to engage with ads differently,
  using engagement as an objective may result in outcomes that reflect these differences.
When an objective that is chosen by the advertiser but is implemented by the platform results in discriminatory delivery, who bears
  the legal responsibility may be unclear.
On one hand, the advertiser (perhaps, unknowingly or implicitly) requested the outcome,
  and if that choice created a discriminatory outcome,
  a prior legal analysis~\cite{datta2018discrimination}
  suggests
  the platform may be protected under Section 230 of the Communications Decency Act,
  a U.S. law that provides ad platforms with immunity from content published by advertisers~\cite{CDA}.
\basi{}
On the other hand, to the extent that a platform might be liable under Title VII,
  one may argue Section 230 does not provide immunity from such liability.
  \comment{prior sen was:
``One possible argument is that the platform should be aware of the risk of skew for job ads when optimizing for conversions,
  and therefore should prevent it, or deny advertisers the option to select this objective, just as they must prevent explicit discriminatory targeting by the advertiser.''
  I suggest next sen instead... the writing seems really unclear with all these ``possible arguments''.  Let's just make the argument and not hide it under indirection.
  ---johnh 2021-04-06}
\basi{}
We suggest that platforms should be aware that ad objectives that optimize for engagement
  may cause delivery algorithms to skew who receives a job ad;
  if it does, the platform may have the responsibility to prevent such skew
     or disable advertiser's choice of such objectives for employment ads
     in order to prevent discrimination.
Our work does not advocate a position on the legal question,
  but provides data (\autoref{sec:reach_results}) about outcomes that shows implications of choices of objectives.

In addition to the optimization objective, other confounding sources of skew ({\autoref{sec:sources_of_skew}})
  may have implications for legal liability.
The prior legal analysis of the Google's ad platform
   evaluated the applicability of Section 230 to different sources of skew,
  and argued Google may not be protected by Section 230 if a skew is
  fully a product of Google's algorithms~\cite{datta2018discrimination}.
Similarly, our goal is to design a methodology
  that controls for confounding factors and isolates skew
  that is enabled solely due to choices made by the platform's ad delivery algorithms.

\section{Background}

We next highlight relevant details about the ad platforms to which we apply
  our methodology and discuss related work.

\subsection{LinkedIn and Facebook Ad Platforms}
  \label{sec:linkedin_background}

We give details about LinkedIn's and Facebook's advertising platforms 
  that are relevant to our methodology.

\textbf{Ad objective:}
LinkedIn and Facebook advertisers purchase ads to meet different marketing \emph{objectives}.
As of February 2021, both LinkedIn and Facebook have three types of objectives: 
  awareness, consideration and conversion, and each type has multiple additional options~\cite{LIAdObjectives, FBAdObjectives}.
For both platforms, the chosen objective constrains the ad format, bidding strategy and
  payment options available to the advertiser.

\textbf{Ad audience:} On both platforms, advertisers can target an audience
  using targeting attributes such as geographic location, age and gender.
But if the advertiser discloses they are running a job ad,
  the platforms disable or limit targeting by age and gender~\cite{FbChanges2020a}.
LinkedIn, being a professional network, also provides targeting by
  job title, education, and job experience.

In addition, advertisers on both platforms can upload a list of known contacts
  to create a custom audience (called ``Matched Audience'' on LinkedIn and ``Custom Audience'' on Facebook).
On LinkedIn, contacts can be specified by first and last name or e-mail address.
Facebook allows specification by many more fields, such as zip code and phone number.
The ad platforms then match the uploaded list
  with profile information from LinkedIn or Facebook accounts.

\textbf{Ad performance report:} Both LinkedIn and Facebook provide
  \emph{ad performance reports} through their website interface and
  via their marketing APIs~\cite{LinkedAPI, FacebookAPI}.
These reports reflect near real-time campaign performance results such as the number of clicks and impressions the ad received,
  broken down along different axes.
The categories of information along which aggregate breakdowns are available differ among platforms.
Facebook reports breaks down performance data by location, age, and gender,
  while LinkedIn gives breakdowns by location, job title, industry and company, but not by age or gender.

\subsection{Related Work}
	\label{sec:related_work}
Targeted advertising has become ubiquitous,
  playing a significant role in shaping information and access to opportunities for hundreds of millions of users.
Because the domains of employment, housing, and credit
  have legal anti-discrimination protections in the U.S.~\cite{DiscriminatoryLaw1, DiscriminatoryLaw2, DiscriminatoryLaw3}, 
the study of ad platform's role in shaping access and exposure to those opportunities has been of particular interest in civil rights discourse~\cite{FbCivilRightsAudit2019, FbCivilRightsAudit2020} and research.
We discuss such work next. 

\textbf{Discriminatory ad targeting:}
Several recent studies consider  
  discrimination in ad targeting:
  \hyphenation{Pro-Pub-li-ca}
journalists at ProPublica were among the first to show that
  Facebook's targeting options enabled job and housing advertisers to discriminate by
  age~\cite{ProPublicaAge}, race~\cite{ProPublicaRace} and gender~\cite{ProPublicaGender}.
In response to these findings and as part of a settlement agreement to a legal challenge~\cite{FacebookEEOCCharges},
Facebook has made changes to 
restrict the targeting capabilities offered to advertisers
  for ads in legally protected domains~\cite{FbChanges2020a, FbChanges2020b}.
Other ad platforms, e.g. Google, have announced similar restrictions~\cite{GoogleChanges2020}.
The question of whether these restrictions are sufficient to stop an ill-intentioned advertiser from discrimination remains open, 
as studies have shown that 
  advanced features of ad platforms,
  such as custom and lookalike audiences,
  can be used to run discriminatory ads~\cite{Speicher2018, sapiezynski2019algorithms, venkatadri-2020-composition, Faizullabhoy2018}. %
Our work assumes a well-intentioned advertiser and performs an audit study
using gender-balanced targeting.

\textbf{Discriminatory ad delivery:}
In addition to the targeting choices by advertisers,
researchers have hypothesized that discriminatory outcomes can be a result of platform-driven choices.
In 2013, Sweeney's empirical study found a statistically significant difference between the likelihood of seeing an ad suggestive of an arrest record on Google when searching for people's names assigned primarily to black babies compared to white babies~\cite{Sweeney2013}.
Datta et al.~\cite{Datta2015} found that the gender of a Google account influences the number of ads one sees related to high-paying jobs, with female accounts seeing fewer such ads.
Both studies could not examine the causes of such outcomes, as their methodology did not have an ability to isolate the role of the platform's algorithm from other possibly contributing factors, such as competition from advertisers and user activity.
Gelauff et al.~\cite{gelauff2020advertising} provide an empirical study of the challenges of advertising to a demographically balanced ad audience without using micro-targeting and in the presence of ad delivery optimization.
Lambrecht et al.~\cite{Lambrecht2016} perform a field test promoting job opportunities in STEM using targeting that was intended to be gender-neutral, find that their ads were shown to more men than women, and explore potential explanations for this outcome.
Finally, recent work by Ali and Sapiezynski et al.~\cite{Ali2019a} has demonstrated that their job and housing ads placed on Facebook are delivered
  skewed by gender and race, even when the advertiser targets a gender- and race- balanced audience,
  and that this skew results from choices of the Facebook's ad delivery algorithm,
  and is not due to market or user interaction effects.
AlgorithmWatch~\cite{AlgWatch2020} replicate these findings %
  with European user audiences, and add an investigation of Google's ad delivery for jobs.
Our work is motivated by these studies,
  confirming results on Facebook and
  performing the first study we are aware of for LinkedIn.
Going a step further to distinguish between skewed and discriminatory delivery, we propose a new methodology to control for user qualifications,
  a factor not accounted for in prior work, but that is critical for evaluating whether skewed delivery is, in fact, discriminatory, for job ads. 
We build on prior work exploring ways in which discrimination may arise in job-related advertising and assessing the legal liability of ad platforms~\cite{datta2018discrimination},
  to establish that
  the job ad delivery algorithms of Facebook may be violating U.S. anti-discrimination law.

\textbf{Auditing algorithms:}
The proprietary nature of ad platforms, algorithms, and their underlying data
  makes it difficult to definitively establish
  the role platforms and their algorithms play for creation of discriminatory outcomes~\cite{Barocas2016, Reisman2018, Bogen2018, Andrus2021, bogen2020awareness}. %
For advertising, in addition to the previously described studies, 
recent efforts have explored the possibility of auditing with data provided by Facebook through its public Ad Library~\cite{FbAdLibrary} (created in response to a legal settlement~\cite{FacebookEEOCCharges}).
Other works have focused on approaches that rely on sock-puppet account creation~\cite{asplund2020auditing, lecuyer2015sunlight}.
Our work uses only ad delivery statistics that platforms
  provide to regular advertisers. 
This approach makes us less reliant on the platform's willingness to be audited. 
We do not rely on transparency-data from platforms,
  since it is often limited and insufficient for answering questions about the platform's role in discrimination~\cite{Mozilla}.
We also do not rely on an ability to create user accounts on the platform,
  since experimental accounts are labor-intensive to create
  and disallowed by most platform's policies.
We build on prior work of external auditing~\cite{Zhang2018, Sandvig2014, Datta2015, datta2018discrimination, Ali2019a, Ali2019b}.
We show that auditing for discrimination in ad delivery of job ads is possible,
  even when limited to capabilities available to a regular advertiser,
  and that one can carefully control for confounding factors.

\textbf{Auditing LinkedIn:} 
To our knowledge, the only work that has studied LinkedIn's ad system's potential for discrimination is that of Venkatadri and Mislove~\cite{venkatadri-2020-composition}.
Their work demonstrates that compositions of multiple targeting options together 
  can result in targeting that is skewed by age and gender,
  without explicitly targeting using those attributes.
They suggest mitigations %
  should be based not on disallowing individual targeting parameters,
  but on the overall outcome of the targeting.
We agree with this goal,
  and go beyond this prior work by basing our evaluation on the outcome of ad delivery, 
 measuring delivery of real-world ads,
  and contrasting outcomes on LinkedIn with Facebook's.

LinkedIn has made efforts to integrate fairness metrics into some of
  its recommendation systems~\cite{Geyik2019, nandy2021achieving}.
Our work looks at a different product, its ad platform, for which,
 to our knowledge, LinkedIn has not made public claims about
  fairness-aware algorithms.

\section{Auditing Methodology}
  \label{sec:methodology}

We next describe the methodology we propose to audit ad delivery algorithms for potential discrimination.

Our approach consists of three steps.
First, we use the advertising platform's custom audience feature (\autoref{sec:audience})
  to build an audience that allows us to infer gender of the ad recipients for platforms
  that do not provide ad delivery statistics along gender lines.
Second,
  we develop a novel methodology that controls for job qualifications
  by carefully selecting job categories (\autoref{sec:job_categories})
  for which everyone in the audience is equally qualified (or not qualified)
  for, yet for which there are distinctions in the real-world gender distributions of employees in the companies.
We then run paired ads concurrently for each job category and
  use statistical tests to evaluate whether the ad delivery results are skewed (\autoref{sec:method_stats}).

Our lack of access to users' profile data, interest or browsing activity prevents us from
  directly testing whether ad delivery satisfies metrics of fairness commonly used in the literature, such as \emph{equality of opportunity}~\cite{Hardt2016}, 
  or recently proposed for ad allocation tasks where users have diverse preferences over outcomes, such as preference-informed individual fairness~\cite{kim2020preference}.
In our context of job ads,
  equality of opportunity means that an individual in a demographic group that is qualified for a job should
  get a positive outcome (in our case: see an ad) at equal rates compared to an equally qualified individual in another demographic group.
While our methodology does not test for this metric,
  we indirectly account for qualification
  in the way we select which job categories we run ads for.

We only describe a methodology for studying discrimination in ad delivery along gender lines,
  but we believe our methodology can be generalized to audit along other attributes such as race and age
  by an auditor with access to auxiliary data that is needed for picking appropriate job categories.

\subsection{Targeted Audience Creation}
	\label{sec:audience}

Unlike Facebook, LinkedIn does not give a gender breakdown of ad impressions,
  but reports their location at the county level.
As a workaround, we rely on an approach introduced in prior work~\cite{Ali2019a, Ali2019b} that uses
 ad recipients' location to infer gender.

To construct our ad audience,
  we use North Carolina's voter record dataset~\cite{VoterData}, which among other fields
  includes each voter's name, zip code, county, gender, race and age.
We divide all the counties in North Carolina into two halves.
We construct our audience by including only male voters from counties in the first half,
  and only female voters from counties in the second half
(this data is limited to a gender binary, so our research follows).
If a person from the first half of the counties is reported as having seen an ad, we can infer that the person is a male,
  and vice versa.
Furthermore, we include a roughly equal number of people from each gender in the targeting
  because we are interested in measuring skew that results from
  the delivery algorithm, not the advertiser's targeting choices.

To evaluate experimental reproducibility without introducing
  test-retest bias,
  we repeat our experiments across different,
  but equivalent audience partitions.
\autoref{tab:audience_summary} gives a summary of the partitions
 we used.
Aud\#0, Aud\#1 and Aud\#2  are partitions whose size is approximately a quarter of the full audience,
  while Aud\#0f, Aud\#1f and Aud\#2f
  are constructed %
  by swapping the choice of gender by county.
Swapping genders this way doubles the number of partitions we can use.

On both LinkedIn and Facebook, the information we upload is used to find
  \emph{exact matches} with information on user profiles.
We upload our audience partitions to LinkedIn in the form of first and last names.
For Facebook, we also include zip codes,
  because their tool for uploading audiences notified us that the match rate would be too low when building audiences only on the basis of first and last names.
The final targeted ad audience is a subset of the audience we upload, because not all the names will be matched, i.e. will correspond to an actual user of a platform.
As shown in~\autoref{tab:audience_summary}, for each audience partition,
  close to 12\% of the uploaded names 
  were matched with accounts on LinkedIn.
Facebook does not report the final match rates for our audiences in order to protect user privacy.

\begin{table}
\caption{Audiences used in our study. }
\label{tab:audience_summary}
\centering
\begin{tabular}{ |c|c|c|c|c| }
\hline 
ID & Size & Males & Females & Match Rate \\ 
\hline 
Aud \#0 & 954,714 & 477,129 & 477,585 & 11.83\% \\ 
Aud \#1 & 900,000 & 450,000 & 450,000 & 11.6\% \\ 
Aud \#2 & 950,000 & 450,000 & 500,000 & 11.8\% \\ 
Aud \#0f & 850,000 & 450,000 & 400,000 & 11.88\% \\ 
Aud \#1f & 800,000 & 400,000 & 400,000 & 12.51\% \\ 
Aud \#2f & 790,768 & 390,768 & 400,000 & 12.39\% \\
\hline
\end{tabular}
\end{table}

To avoid self-interference between our ads over the same audience
  we run paired ads concurrently,
  but ads for different job categories
  or for different objectives sequentially.
In addition, to avoid test-retest bias,
  where a platform learns from prior experiments who is likely to respond
  and applies that to subsequent experiments,
  we generally use different (but equivalent) target audiences.
  
\subsection{Controlling for Qualification}
  \label{sec:job_categories}

The main goal of our methodology is to distinguish
  skew resulting from algorithmic choices that are not related to qualifications,
from skew that can be justified by differences in user qualifications for the jobs advertised.
A novel aspect of our methodology is to
  \emph{control} for qualifications
  by running paired ads for jobs %
  with similar qualification requirements,
  but skewed de facto gender distributions.
We measure skew by comparing the
  \emph{relative difference} between the delivery of
  a \emph{pair of ads} that run concurrently, targeting the same audience.
Each test uses paired jobs that meet two criteria:
First, they must have \emph{similar qualification requirements},
  thus ensuring that the people that we
  target our ads with are equally qualified (or not qualified) for both job ads.
Second, the jobs must exhibit a \emph{skewed, de facto
  gender distribution} in the real-world,
  as shown through auxiliary data.
Since both jobs require similar qualifications,
  our assumption is that on a platform whose ad delivery algorithms are non-discriminatory,
  the distribution of genders among the recipients of the two ads will be roughly equal.
On the other hand,
in order to optimize for engagement or business objectives,
  platforms may incorporate other factors into ad delivery optimization,
  such as training or historical data.
This data may reflect the de facto skew and thus influence machine-learning-based algorithmic predictions of engagement.
Since such factors do not reflect differences in job qualifications,
  they may be disallowed (\autoref{sec:liability})
  and therefore represent platform-induced discrimination
  (even if they benefit engagement or
  the platform's business interests).
We will look for evidence of such factors in a \emph{difference} in gender distribution
  between the paired ads (see~\autoref{sec:skew} for how we quantify the difference).

In~\autoref{sec:skew_results}, we use the above criteria to select three job
  categories -- delivery driver, sales associate and software engineer -- and
  run a pair of ads for each category and compare the gender make-up
  of the people to whom LinkedIn and Facebook show our ads.
An example of such a pair of ads is a delivery driver job at Domino's
  (a pizza chain) and at Instacart (a grocery delivery service).
The de facto gender distribution among drivers of these services is skewed male for Domino's
  and skewed female for Instacart~\cite{DominosGender, InstacartGender}.
If a platform shows the Instacart ad to relatively more women than a Domino's ad, 
  we conclude that the platform's algorithm is discriminatory, since both jobs have similar qualification requirements and thus a gender skew cannot be attributed to differences in qualifications across genders represented in the audience.

Using paired, concurrent ads that target the same
  audience also ensures other confounding factors such as timing or competition from other advertisers affect both ads equally~\cite{Ali2019a}.

To avoid bias due to the audience's willingness to move for a job,
  we select jobs in the same physical location.
When possible (for delivery driver and sales job categories, but not software engineering),
  we select jobs in the location of our target audience.

\subsection{Placing Ads and Collecting Results}
	\label{sec:method_stats}

We next describe the mechanics of placing ads on Facebook and LinkedIn, and collecting
  the ad delivery statistics which we use to calculate the gender breakdown
  of the audiences our ads were shown to.
We also discuss the content and parameters we use for running our ads.

\subsubsection{Ad Content}
  \label{sec:ad_content}

\begin{figure}
\begin{minipage}[b]{0.48\linewidth}
\centering
\includegraphics[width=\textwidth]{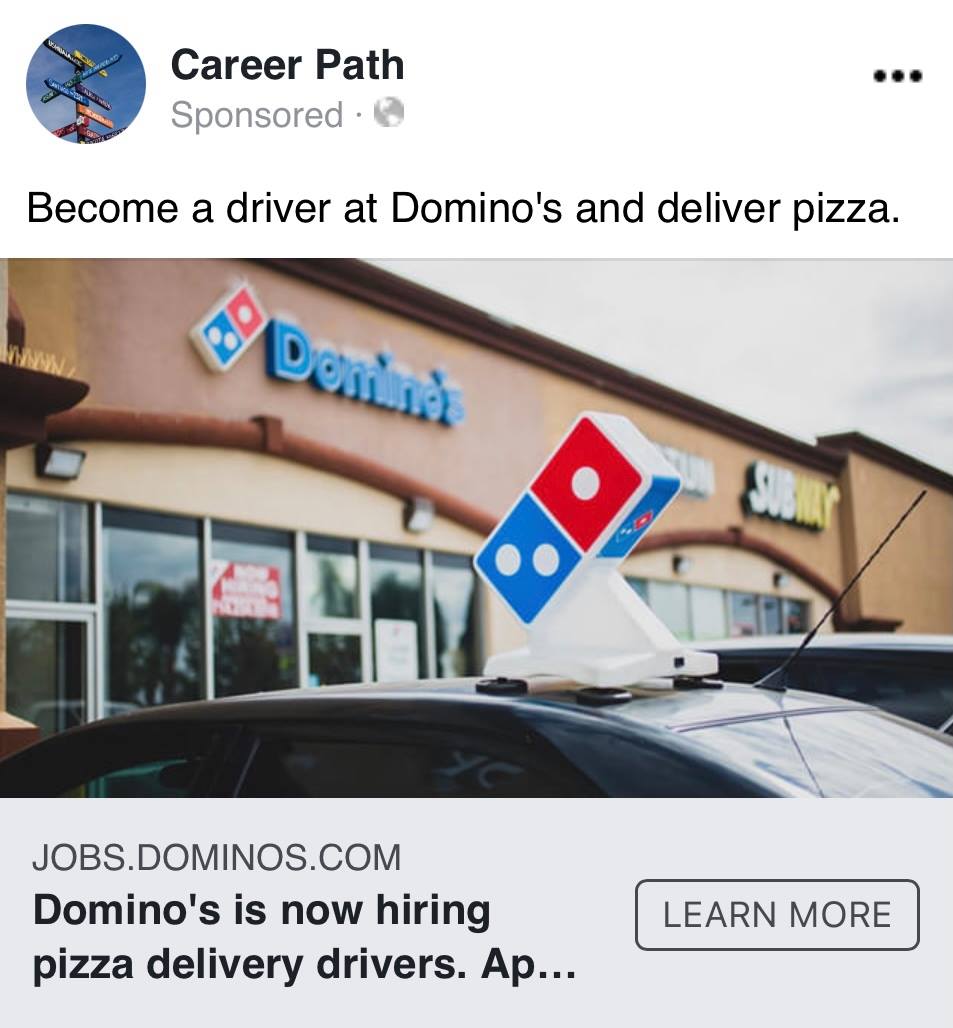}
\end{minipage}
\hspace{0.1cm}
\begin{minipage}[b]{0.48\linewidth}
\centering
\includegraphics[width=\textwidth]{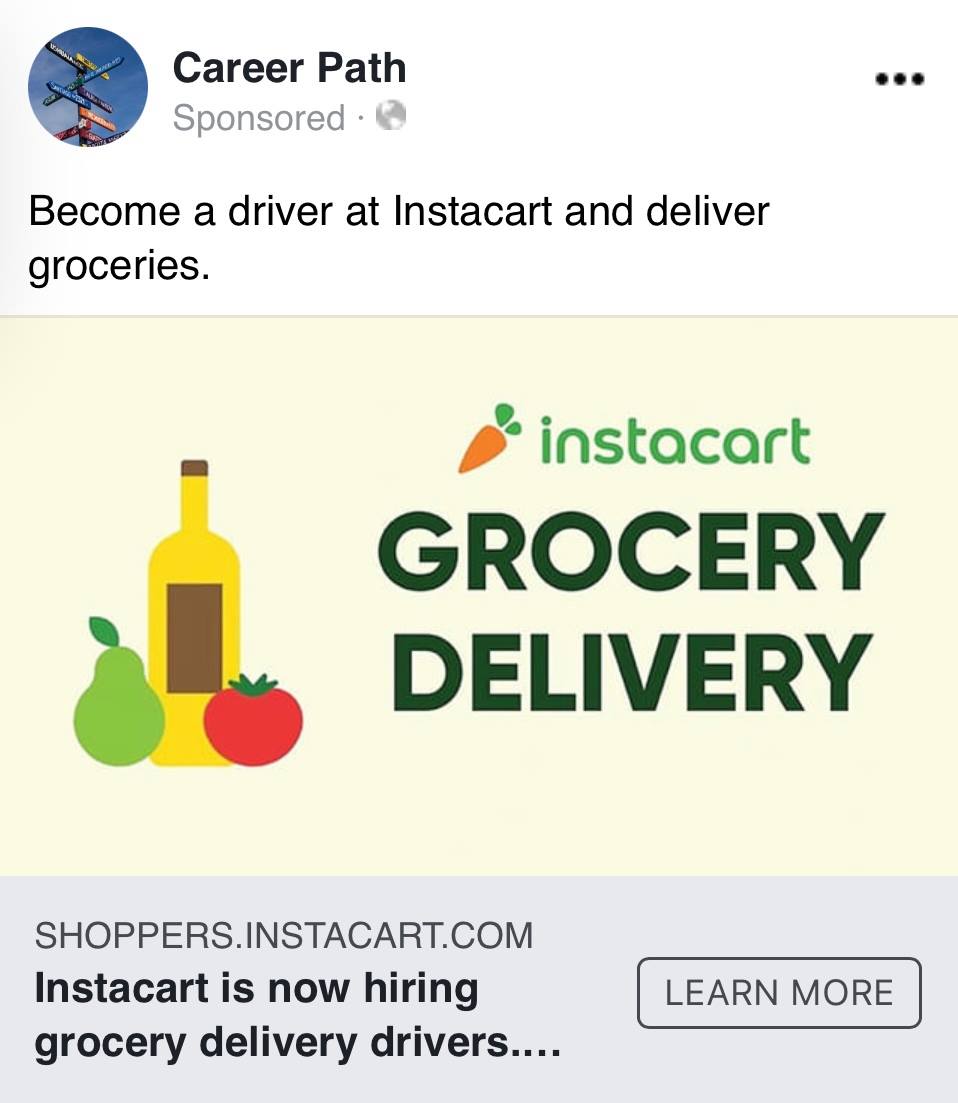}
\end{minipage}
\caption{Example delivery driver job ads
  for Domino's and Instacart.}
\label{fig:ad_screenshots}
\end{figure}

In creating our ads, we aim to use gender-neutral text and image so as to minimize any possible skew due to the input of an advertiser (us).
The ad headline and description for each pair of ads is customized to each job category as described in~\autoref{sec:skew_results}.
Each ad we run links to a real-world job opportunity that
  is listed on a job search site,
  pointing to a job posting on a company's careers page (for delivery driver) or to a job posting on LinkedIn.com (in other cases).
\autoref{fig:ad_screenshots} shows screenshots
  of two ads from our experiments.

\subsubsection{Ad Optimization Objective}\label{sec:objective}
We begin by using the \emph{conversion} objective because searching for people who are likely to take an action on the job ad
  is a likely choice for advertisers seeking users who will
  apply for their job (\autoref{sec:skew_results}).
For LinkedIn ads, we use ``Job Applicants'' option,
  a conversion objective with the goal:
 ``Your ads will be shown to those most likely to view or click on your job ads, getting more applicants.''~\cite{LIAdObjectives}.
For Facebook ads, we use ``Conversions'' option with
  with the following optimization goal:
  ``Encourage people to take a specific action on your business's site''~\cite{FBAdObjectives},
  such as register on the site or submit a job application.

In~\autoref{sec:reach_results}, we run some of our Facebook ads using the \emph{awareness} objective.
By comparing the outcomes across the two objectives we can evaluate whether an advertiser's objective choice plays a role in the skew (\autoref{sec:liability}).
We use the ``Reach'' option that Facebook provides 
  within the awareness objective with the stated goal of:
  ``Show your ad to as many people as possible in your target audience''~\cite{FBAdObjectives}.

\subsubsection{Other Campaign Parameters}
	\label{sec:other_parameters}

We next list other parameters we use for running ads and our reasons for picking them.

From the ad format options available for the objectives we selected, we choose
  \emph{single image ads}, which show up in a prominent part of LinkedIn and Facebook users' newsfeeds.

We run all Facebook and LinkedIn ads with a total budget of \$50 per ad campaign and schedule them to run
  for a full day or until the full budget is exhausted.
This price point ensures
  a reasonable sample size for statistical evaluations,
  with all of our ads receiving at least 340 impressions.

For both platforms, we request automated bidding to maximize the number of clicks (for the conversion objective) and impressions (for the awareness objective)
  our ads can get within the budget.
We configure our campaigns on both platforms to pay per impression shown.
On LinkedIn, this is the only available option for our chosen parameters.
We use the same option on Facebook for consistency.
On both platforms we disable audience expansion and off-site delivery options.
While these options might show our ad to more users,
  they are not relevant or may interfere with our methodology.

Since our methodology for LinkedIn relies on using North Carolina county names as proxies
  for gender, we add ``North Carolina'' as the location for our target audience.
We do the same for Facebook for consistency across experiments but
  we do not need to use location as a proxy to infer gender in Facebook's case. 

\subsubsection{Launching Ads and Collecting Delivery Statistics}

For LinkedIn, we use its \emph{Marketing Developer Platform} API to create the ads,
  and once the ads run, to get the final count of impressions per county which we use
  to infer gender.
For Facebook, we create ads via its Ad Manager portal.
The portal gives a breakdown of ad impressions by gender, so we do not
  rely on using county names as a proxy.
We export the final gender breakdown after the ad completes running.

\subsection{Skew Metric}
  \label{sec:skew}

We now describe the metric we apply to the \emph{outcome} of advertising, i.e. the demographic make-up of the audience that saw our ads, to establish whether platform's ad delivery algorithm leads to discriminatory outcomes.  

\subsubsection{Metric:}
\begin{figure}
  \begin{minipage}[c]{0.3\columnwidth}
    \includegraphics[width=\columnwidth]{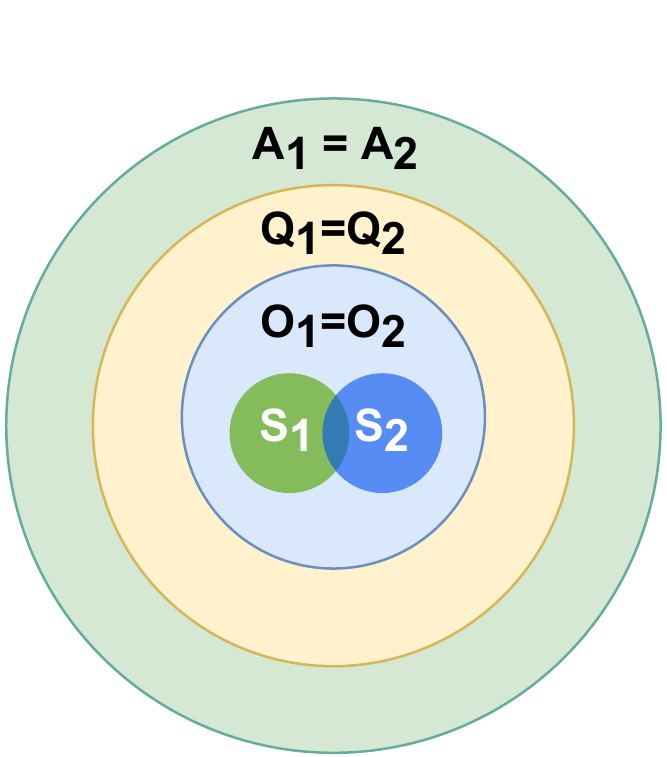}
  \end{minipage}\hfill
  \begin{minipage}[c]{0.6\columnwidth}
    \caption{
     Relation between subsets of audiences involved in running
     two ads targeting the same audience.
     The subscripts indicate sets for the first and second ad.
    } \label{fig:ad_delivery_venn}
  \end{minipage}
\end{figure}

As discussed in the beginning of this section,
  out methodology works by running two ads simultaneously and looking at the relative difference in
  how they are delivered.
In order to be able to effectively compare delivery of the two ads,
  we need to ensure the baseline audience that we use to measure skew is the same
  for both ads.
The baseline we use is people who are qualified for the
  job we are advertising and are browsing the platform during the ad campaigns.
However, we must consider several audience subsets shown in \autoref{fig:ad_delivery_venn}:
$A$, the the audience targeted by us, the advertiser (us);
$Q$, the subset of $A$ that the ad platform's algorithm considers qualified for the
  job being advertised,
and $O$, the subset of $Q$ that are online when the ads are run.

Our experiment design should ensure
  that these sets are the same for both ads, so that
  a possible skewed delivery cannot be merely explained by a difference
  the underlying factors
  these sets represent. 
We ensure $A$, $Q$, and $O$ match for our jobs 
  by targeting the same audience (same $A$),
  ensuring both jobs have similar qualification requirements (same $Q$) as discussed in~\autoref{sec:job_categories},
  and by running the two ads at the same time (same $O$).

To \emph{measure gender skew}, we compare what fraction of people in
  $O$ that saw our two ads are a member of a specific gender.
Possible unequal distribution of gender in the audience does not affect our comparison
  because it affects both ads equally (because $O$ is the same for both ads).
Let $S_1$ and $S_2$ denote subsets of people in $O$ who saw the first and second ad,
  respectively.
$S_1$ and $S_2$ are not necessarily disjoint sets.
To measure gender skew, we compare the fraction of females in $S_1$ that saw the
  first ad ($s_{1,f}$) and fraction of females in $S_2$ that saw the second ad ($s_{2,f}$)
  with the fraction of females in $O$ that were online during the ad campaign ($o_f$).

In the absence of discriminatory delivery,
  we expect, for both ads, the gender make-up of the audience the ad is shown to be
  representative of the gender make-up of people that were online and participated
  in ad auctions. Mathematically, we expect $s_{1,f} = o_f$ and $s_{2,f} = o_f$.
As an external auditor that does not have access to users' browsing activities,
  we do not have a handle on $o_f$ but we can directly compare
  $s_{1,f}$ and $s_{2,f}$.
Because we ensure other factors that may affect ad delivery are either controlled or affect both ads equally,
  we can attribute any difference we might observe between $s_{1,f}$ and $s_{2,f}$
  to choices made by the platform's ad delivery algorithm based on factors unrelated to qualification of users,
  such as revenue or engagement goals of the platform.

\subsubsection{Statistical Significance:}
  \label{sec:stat_sig}

We use the Z-Test to
  measure the statistical significance of a difference in proportions we observe 
  between $s_{1,f}$ and  $s_{2,f}$.
Our null hypothesis is that there is no gender-wise difference between the audiences
  that saw the two ads, i.e., $s_{1,f} = s_{2,f}$,
  evaluated as:
$$
Z = \frac{s_{1,f} - s_{2,f}}{\sqrt{ \hat{s}_f ( 1 - \hat{s}_f ) (\frac{1}{n_1} + \frac{1}{n_2}) }}
$$
where $\hat{s}_f$ is fraction of females in $S_1$ and $S_2$ combined ($S_1 \cup S_2$),
  and $n_1$ and $n_2$ are the sizes of $S_1$ and $S_2$, respectively.
At $\alpha$ significance level, if $Z > Z_\alpha$, we reject the null hypothesis and conclude
  that there is a statistically significant gender skew in the ad delivery.
We use a 95\% confidence level ($Z_\alpha = 1.96$) for all of our statistical tests.
This test assumes the samples are independent and $n$ is large.
Only the platform knows whom it delivers the ad to,
  so only it can verify independence.
Sample sizes vary by experiment, as shown in figures,
  but they always exceed 340 and often are several thousands.

\subsection{Ethics}

Our experiments are designed to consider ethical implications,
  minimizing harm both to the platforms and the individuals that interact with our ads.
We minimize harm to the platforms by registering as an advertiser and
  interacting with the platform just like any other regular advertiser would.
We follow their terms of service,
  use standard APIs available to any advertiser and do not collect any user data.
We minimize harm to individuals using the platform and seeing our ads
  by having all our ads link to a real job opportunity as described.
Finally, our ad audiences aim to include an approximately
  equal number of males and females and so aim not to discriminate.
Our study was classified as exempt by our Institutional Review Board.

\section{Experiments}
  \label{sec:experiments}

We next present the results from applying our methodology to real-world ads on Facebook and
  LinkedIn.
We find contrasting results that show statistically significant evidence of skew that is not justifiable
  on the basis of qualification in the case of Facebook, but not in the case of LinkedIn. 
  We make data for the ads we used in our experiments and their delivery statistics
  publicly available at~\cite{AdDeliveryDataset}.
We ran all ads in February, 2021.

\subsection{Measuring Skew in Real-world Ads}
  \label{sec:skew_results}

We follow the criteria discussed in~\autoref{sec:job_categories} to pick and compare
  jobs which have similar qualification requirements but for which there is data that shows
  the de facto gender distribution is skewed.
We study whether ad delivery optimization algorithms reproduce these de facto skews,
  even though they are not justifiable on the basis of differences in qualification.

We pick three job categories: a low-skilled job (delivery driver),
  a high-skilled job (software engineer),
  and a low-skilled but popular job among our ad audience (sales associate).
Since our methodology compares two ads for each category,
  we select two job openings at companies for which we have evidence of de facto gender distribution differences,
  and use our metric~\autoref{sec:skew} to measure whether there is a statistically
  significant gender skew in ad delivery.
In each job category,
  we select pairs of jobs in the same state
  to avoid skew (\autoref{sec:job_categories}).

For each experiment, we run the same pair of ads on both Facebook and LinkedIn and compare
  their delivery.
For both platforms, 
  we repeat the experiments on three different audience partitions
  for reproducibility.
We run the ads for each job category at different times to avoid self-competition
  (\autoref{sec:audience}).
We run these first set of ads using the conversion objective~(\autoref{sec:objective}).

As discussed in~\autoref{sec:ad_content}, we build our ad creatives (text and image) using
  gender-neutral content to minimize any skew due to an advertiser's (our)
  input.
For delivery driver and sales associate categories,
  Facebook ad text uses modified snippets of the real job descriptions they link to
    (for example,
    ``Become a driver at Domino's and deliver pizza'').
Images use a company's logo or a picture of its office.
To ensure any potential skew is not due to keywords in the job descriptions that
  could appeal differently to different audiences,
  we ran the software engineering Facebook ads using generic headlines
  with a format similar to the ones shown in~\autoref{fig:ad_screenshots}, and found similar results to the runs that used modified snippets.
All LinkedIn ads were ran using generic ad headlines similar to those in~\autoref{fig:ad_screenshots}.

\begin{figure}
  \centering
  \begin{subfigure}{\columnwidth}
    
    \includegraphics[width=1\columnwidth]{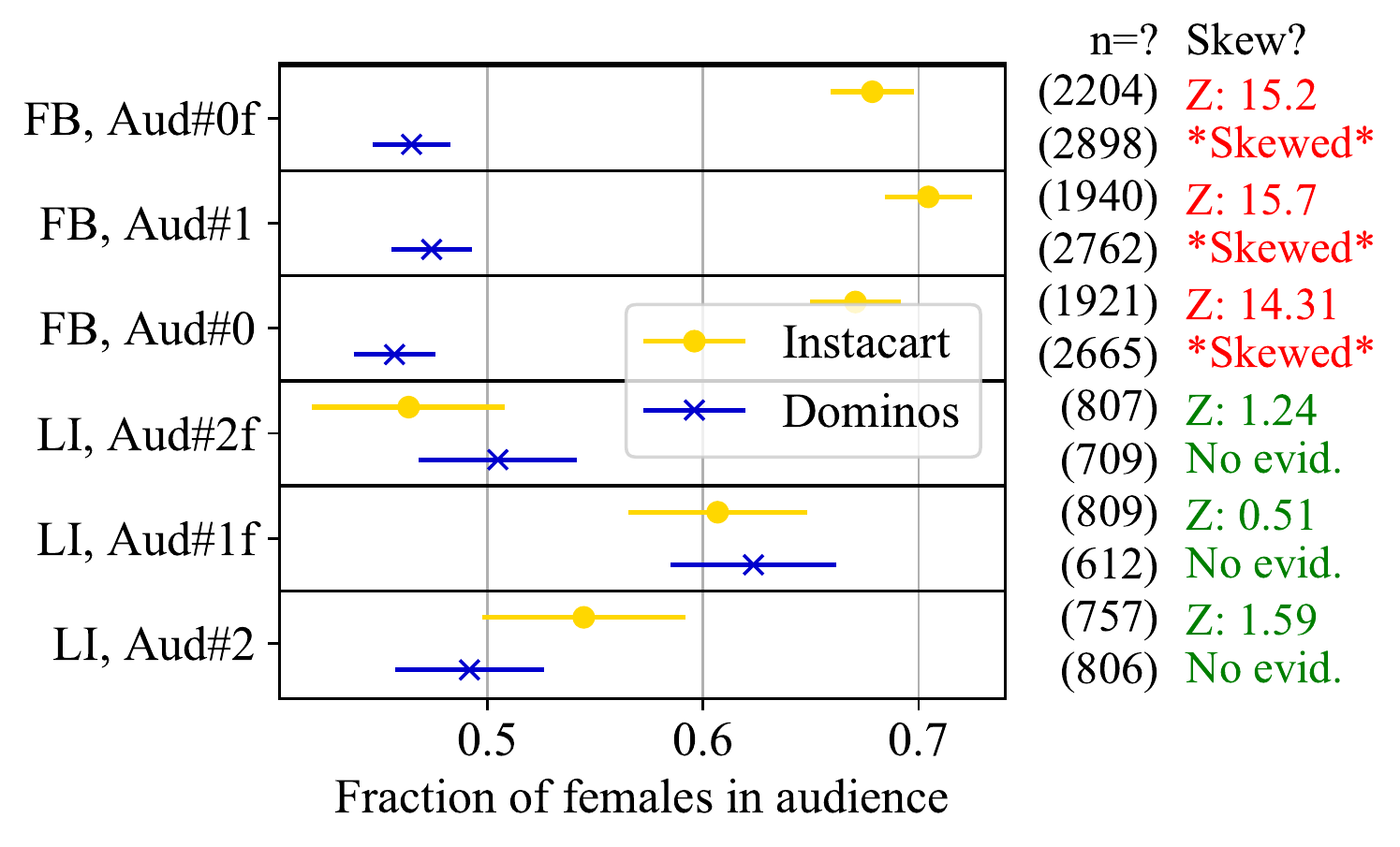}
    \caption{Delivery Driver at Domino's vs. Instacart}
    \label{fig:delivery_driver}
   
  \end{subfigure}
  \begin{subfigure}{\columnwidth}
    
    \includegraphics[width=1\columnwidth]{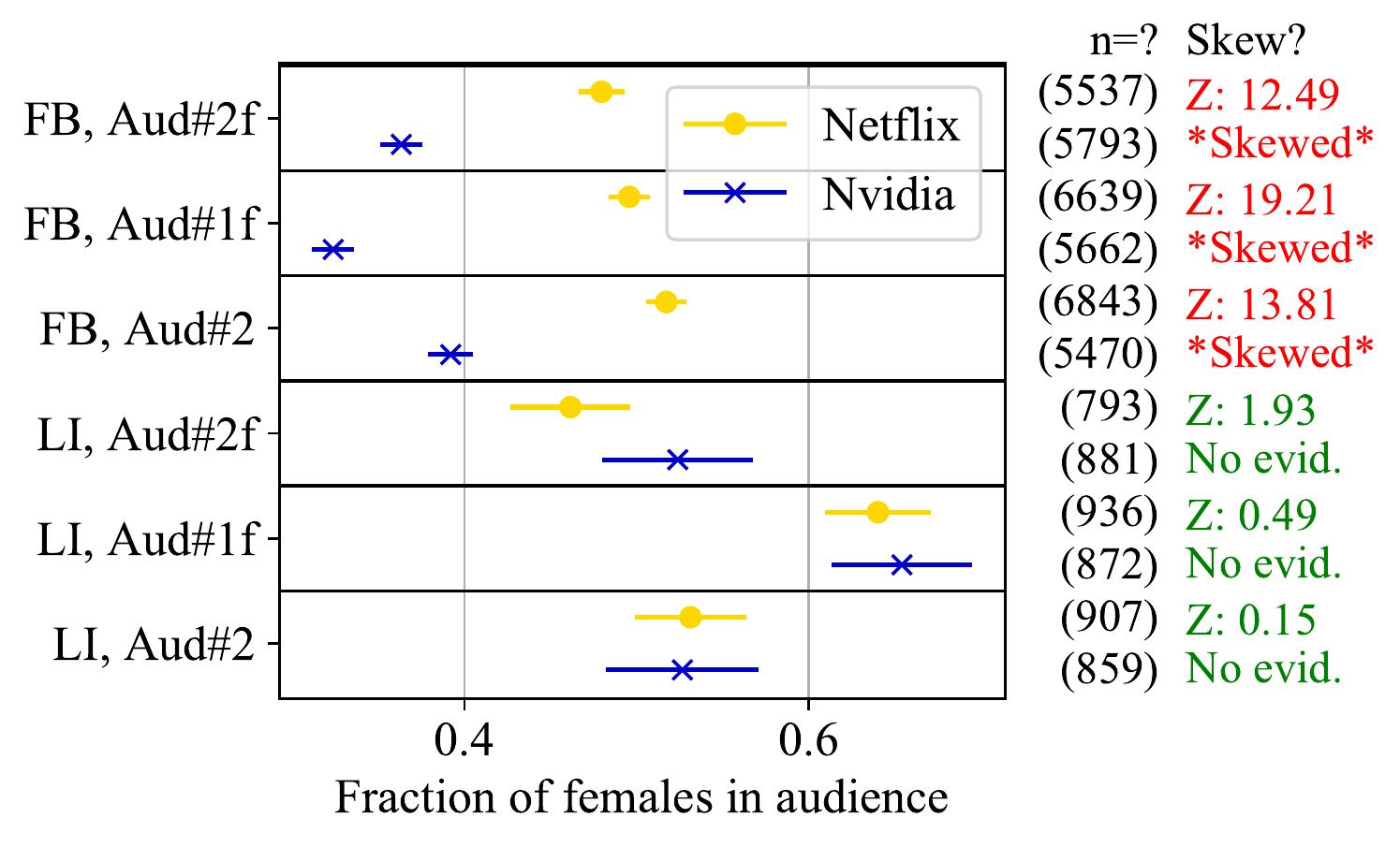}
    \caption{Software Engineer at Nvidia vs. Netflix}
    \label{fig:software_engineer}
    
  \end{subfigure}
  \begin{subfigure}{\columnwidth}
    \includegraphics[width=1\columnwidth]{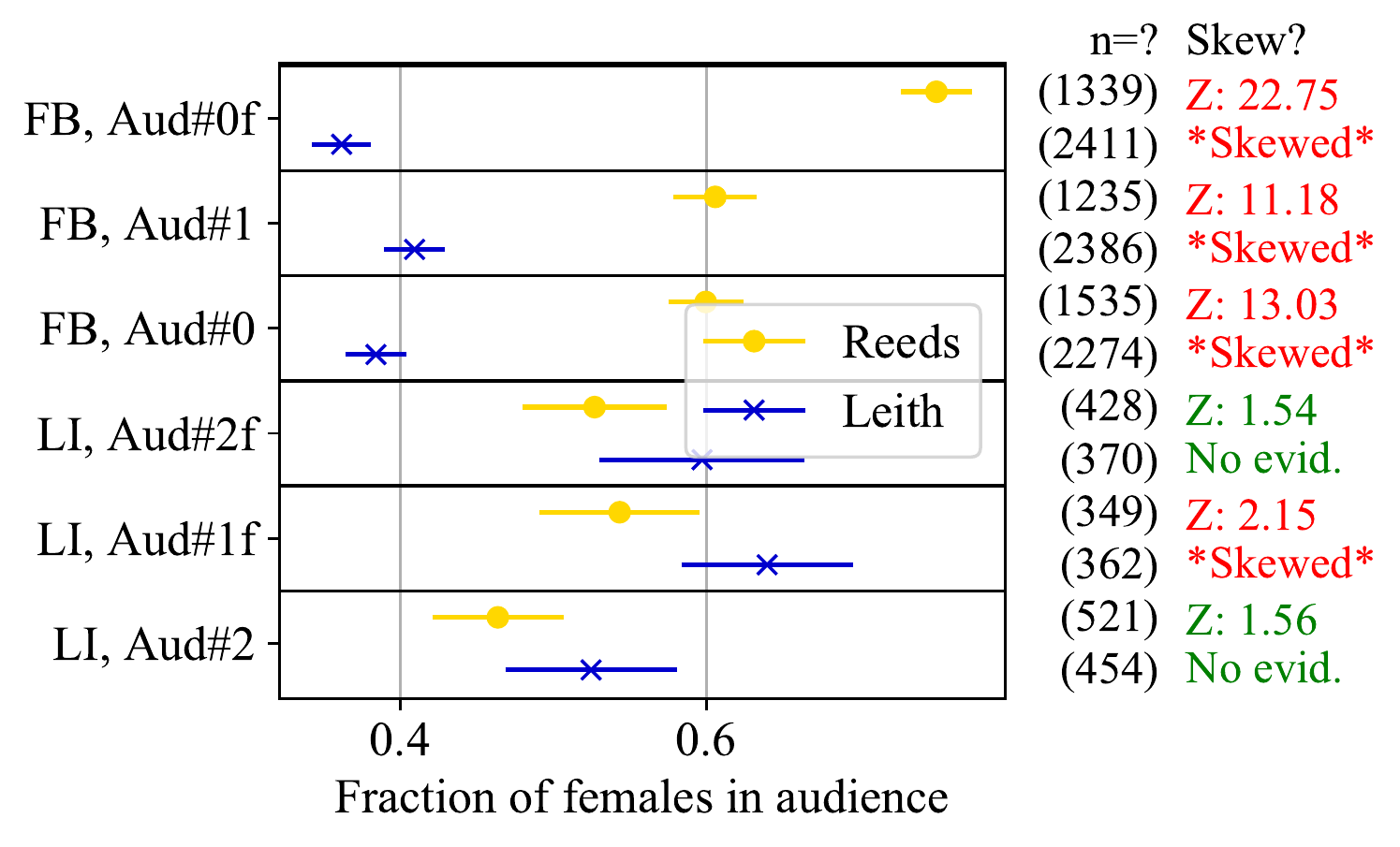}
    \caption{Sales Associate at Reeds Jewelry vs. Leith Automotive}
    \label{fig:sales_associate}
    
  \end{subfigure}
  \caption{Skew in delivery of real-world ads on Facebook (FB) and LinkedIn (LI), using ``Conversion'' objective. $n$ gives total number of impressions.
  We use our metric (\autoref{sec:skew}) to test for skew at 95\% confidence level ($Z > 1.96$).}
  \label{fig:skew_results}
\end{figure}

\subsubsection{Delivery Drivers}
We choose \emph{delivery driver} as a job category to study
  because we were able to identify two companies -- Domino's and Instacart -- with significantly
  different de facto gender distributions among drivers,
  even though their job requirements are similar.
98\% of delivery drivers for Domino's are male~\cite{DominosGender}, 
  whereas more than 50\% of Instacart drivers are female~\cite{InstacartGender}.
We run ads for driver positions in North Carolina for both companies,
  and expect a platform whose ad delivery optimization 
    goes beyond what is justifiable by qualification and 
  reproduces de facto skews
  to show the Domino's ad to relatively more males than the Instacart ad.

\autoref{fig:delivery_driver} shows
  gender skews in the results of ad runs for delivery drivers,
  giving the gender ratios of ad impressions with 95\% confidence intervals.
These results
  show evidence of a \emph{statistically significant
  gender skew on Facebook}, and show \emph{no gender skew on LinkedIn}.
The skew we observe on Facebook is in the same direction as the de facto skew, 
  with the Domino's ad delivered to a higher fraction of men than the Instacart ad.
We confirm the results across three separate runs for both platforms,
  each time targeting a different audience partition.

\subsubsection{Software Engineers}\label{sec:swe}
We next consider the \emph{software engineer} (SWE) job category, a high-skilled job
    which may be a better match for LinkedIn users than delivery driver jobs.

We pick two companies based employee demographics stated in their diversity report .
Because we are running software engineering ads, we specifically look at
  the percentage of female employees who work in a tech-related position.
We pick Netflix and Nvidia for our paired ad experiments.
At Netflix, 35\% of employees in tech-related positions are female~\cite{NetflixGender} according to its 2021 report.
At Nvidia, 19\% of all employees are female according to~\cite{NvidiaGender}, 
and third-party data as of 2020 suggests that the percentage of female employees in tech-related positions is as low as 14\%~\cite{NvidiaGender2}.
For both companies, we find job openings in the San Francisco Area and run
  ads for those positions. 
We expect a platform whose algorithm learns and perpetuates the existing difference in employee demographics
  will show the Netflix ad to more women than the Nvidia ad.

\autoref{fig:software_engineer} shows the results.
The \emph{Facebook results show skew by gender in all three trials},
  with a statistically different gender distribution between the delivery of the two ads.
The skew is in the direction that confirms our hypothesis,
  a higher fraction of women seeing the Netflix ads than the Nvidia ads.
\emph{LinkedIn results are not skewed in all three trials}.
These results confirm the presence of delivery skew not justified by qualifications on Facebook
  for a second, higher-skilled job category.

\subsubsection{Sales Associates}
We consider \emph{sales associate} as a third job category.
Using LinkedIn's audience estimation feature,
  we found that many LinkedIn users in the audience we use identified as having sales experience,
  so we believe people with experience in sales are well-represented in the audience.
The Bureau of Labor Statistics (BLS) data
  shows that sales jobs skew by gender in different industries,
    with women filling 62\% of sales associates in jewelry stores
    and only 17.9\% in auto dealerships~\cite{BLSData}.
We pick Reeds Jewelers~(a retail jeweler) and Leith Automotive
  (an auto dealership) to represent these two industries
  with open sales positions in North Carolina.
If LinkedIn's or Facebook's delivery mimics skew in the de facto
  gender distribution,
  we expect them to deliver the Reeds ad to relatively more women than the Leith ad.

\autoref{fig:sales_associate} presents the results.
All three trials on both platforms confirm our prior results using other job categories,
  with \emph{statistically significant delivery skew between all jobs on Facebook
  but not for two of the three cases on LinkedIn}.
One of the three trials on LinkedIn (Aud\#1f)
  shows skew just above the threshold for a statistical significance,
  and surprisingly it shows bias in the opposite direction from expected
  (more women for the Reeds ad).
We observe that these cases show the smallest response rates (349 to 521)
  and their Z-scores (1.54 to 2.15) are close to the threshold ($Z=1.96$),
  while Facebook shows consistently large skew (11 or more).

\subsubsection{Summary:}
These experiments confirm that our methodology proposed in~\autoref{sec:job_categories} is feasible to implement in practice.
Moreover, the observed outcomes are different among the two platforms.
Facebook's job ad delivery is skewed by gender, even when the advertiser is targeting a gender-balanced audience, consistent with prior results of~\cite{Ali2019a}.
However, because our methodology controls for qualifications, our results imply that the skew cannot be explained by the ad delivery algorithm merely reflecting differences in qualifications.
\basi{}
Thus, based on the discussion of legal liability in~\autoref{sec:liability},
  our findings suggests that Facebook's algorithms may be responsible for
  unlawful discriminatory outcomes.

Our work provides the first analysis of LinkedIn's ad delivery algorithm.
With the exception of one experiment, we did not find evidence of skew by gender introduced by LinkedIn's ad delivery,
  a negative result for our investigation,
  but perhaps a positive result for society.

\subsection{``Reach'' vs.~``Conversion" Objectives}
  \label{sec:reach_results}

In~\autoref{sec:skew_results}, we used the \emph{conversion} objective,
  assuming that this objective would be chosen by most employers running ads and aiming to maximize the number of job applicants.
However, both LinkedIn and Facebook also offer advertisers the choice of the \emph{reach} objective,
  aiming to increase the number of people reached with (or shown) the ad, rather than the number of people who apply for the job.
We next examine how the use of the \emph{reach} objective affects skew in ad delivery on Facebook, compared to the use of the
 \emph{conversion} objective.
We focus on Facebook because we observed evidence of skew that cannot be explained by differences in qualifications in their case,
  and we are interested in exploring whether that skew remains even with a more ``neutral" objective.
While there may be a debate about allocating responsibility for discrimination between advertiser and platform
  when using a conversion objective (see \autoref{sec:liability}),
  we believe that the responsibility for any discrimination observed when the advertiser-selected objective is \emph{reach} rests on the platform.

We follow our prior approach (\autoref{sec:skew_results})
  with one change: we use \emph{reach} as an objective
  and compare with the prior results that used the \emph{conversion} objective.
The job categories and other parameters remain the same and we repeat the experiments on different
  audience partitions for reproducibility.

\begin{figure}
  \centering
  \begin{subfigure}{\columnwidth}
    
    \includegraphics[width=1\columnwidth]{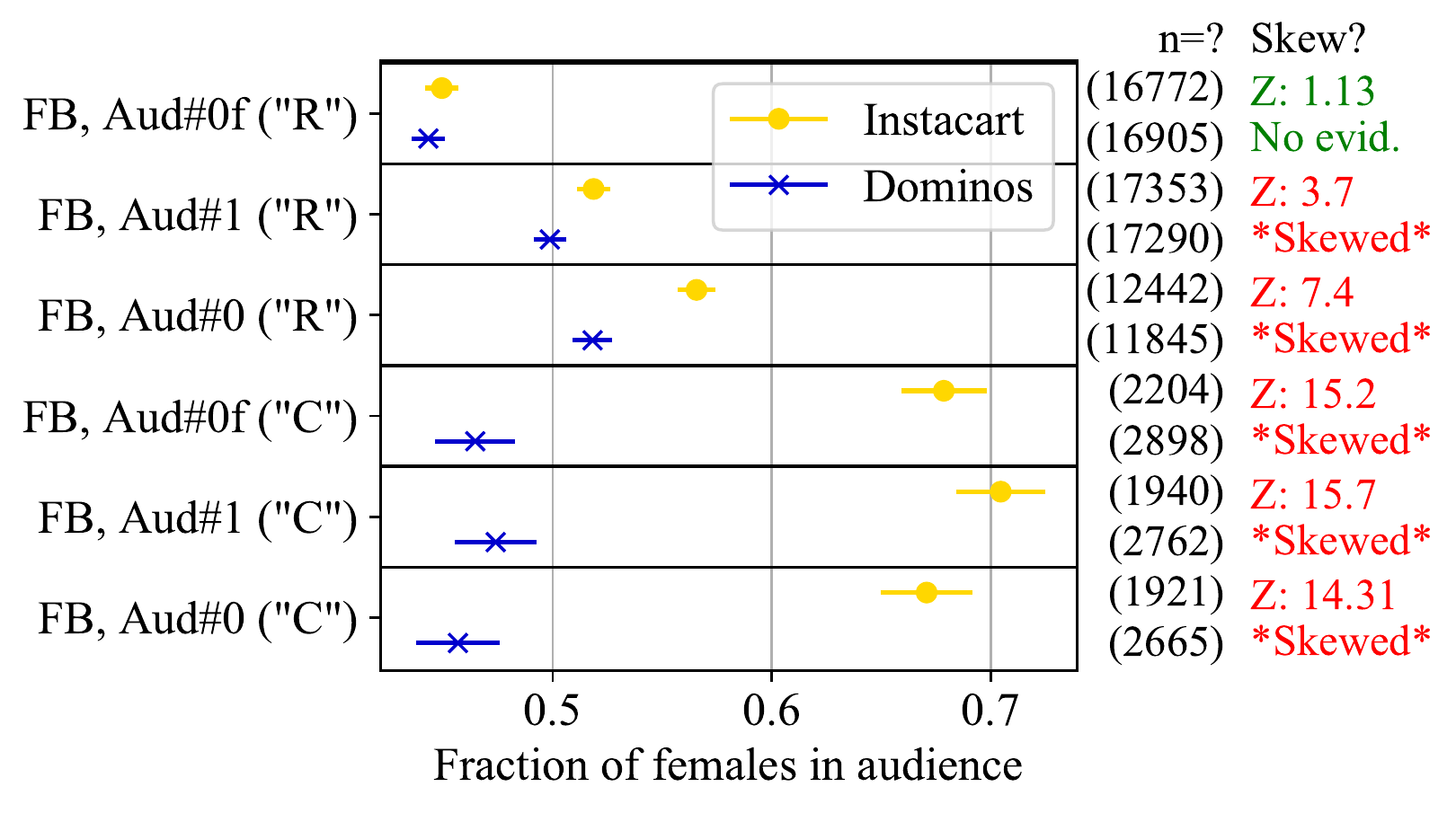}
    \caption{Delivery Driver at Domino's vs. Instacart}
    \label{fig:reach_results_delivery}
   
  \end{subfigure}
  \begin{subfigure}{\columnwidth}
    
    \includegraphics[width=1\columnwidth]{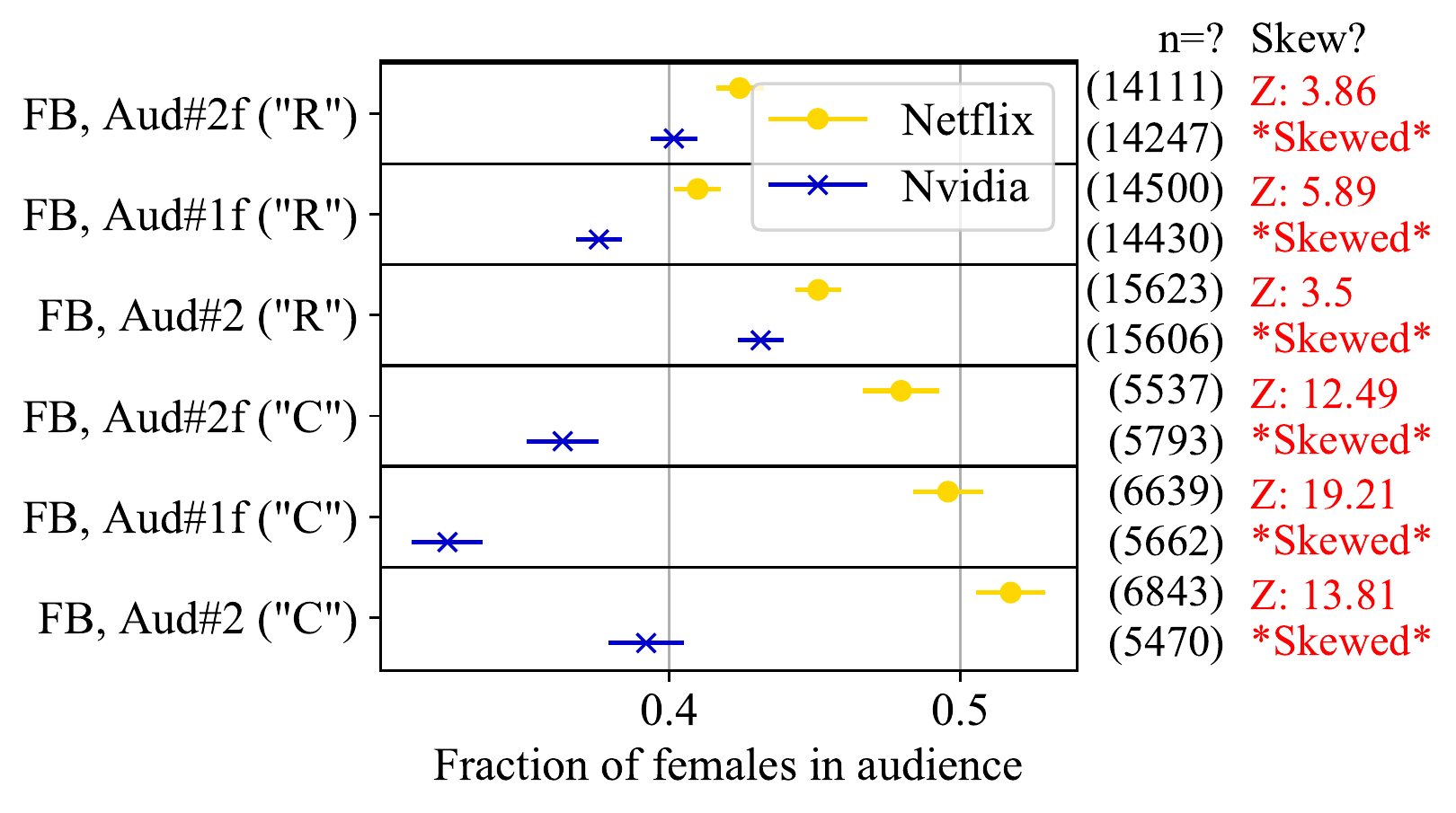}
    \caption{Software Engineer at Nvidia vs. Netflix}
    \label{fig:reach_results_swe}
    
  \end{subfigure}
  \begin{subfigure}{\columnwidth}
    \includegraphics[width=1\columnwidth]{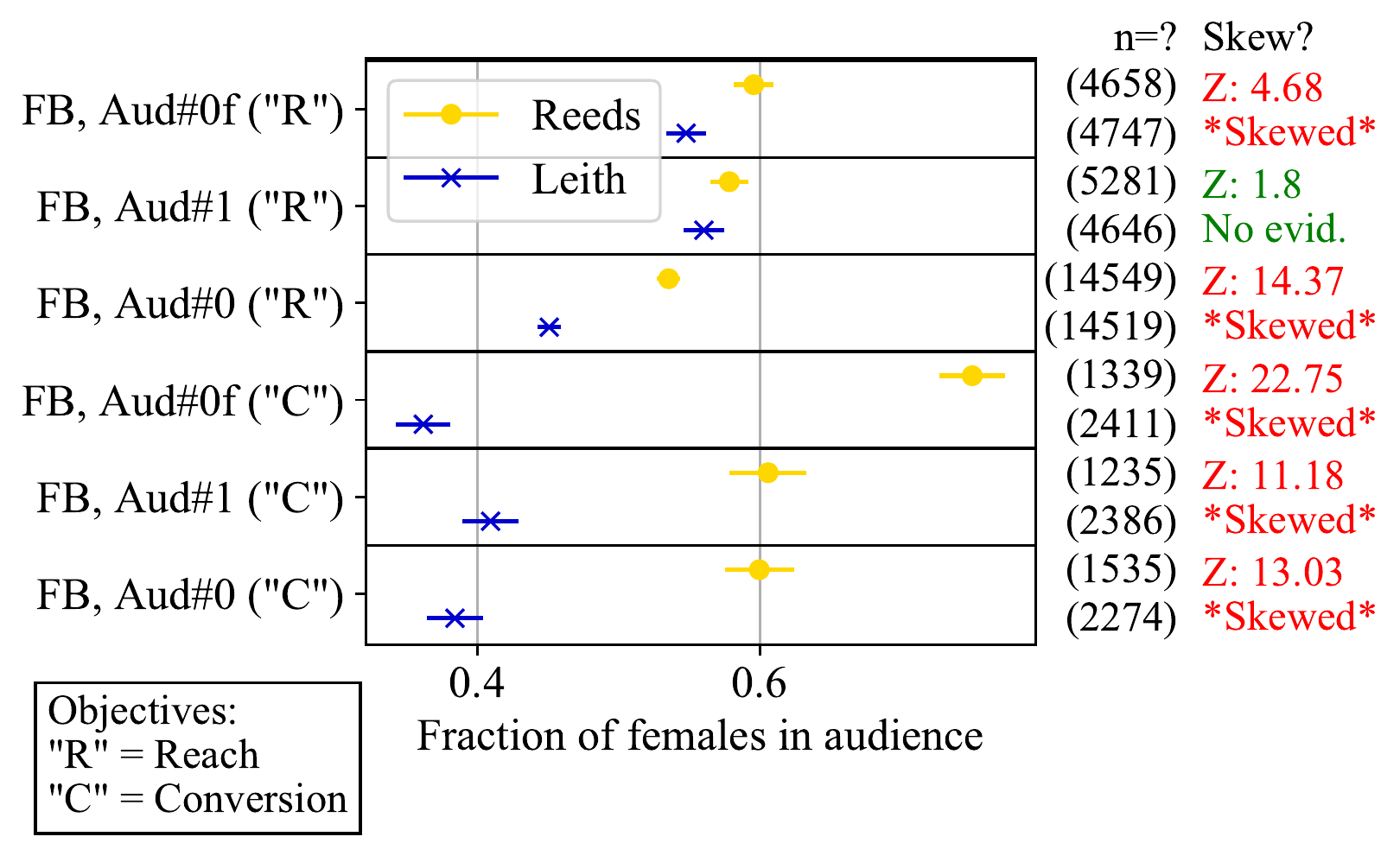}
    \caption{Sales Associate at Reeds Jewelry vs. Leith Automotive}
    \label{fig:reach_results_sales}
    
  \end{subfigure}
  \caption{Comparison of ad delivery with ``reach'' and ``conversion'' objectives on Facebook.}
  \label{fig:reach_results}
\end{figure}

\autoref{fig:reach_results_delivery}, \autoref{fig:reach_results_swe} and \autoref{fig:reach_results_sales}
  show the delivery of \emph{reach} ads for the delivery driver, software engineer and sales associate jobs, respectively.
For comparison, the figures include the prior Facebook experiments ran using \emph{conversion} objective (from \autoref{fig:skew_results}).
For all three job categories,
  the results show a statistically significant skew in at least two out of the three experiments using the reach objective.
This result confirms our result in~\autoref{sec:skew_results} that showed Facebook's ad delivery algorithm
  introduces gender skew even when advertiser targets a gender-balanced audience.
Since skewed delivery occurs even when the advertiser chooses the \emph{reach} objective,
  the skew is attributable to the platform's algorithmic choices and not to the advertiser's choice.

On the other hand, we notice two main differences in the delivery of the ads run with the \emph{reach} objective.
For all three job categories (\autoref{fig:reach_results_delivery}, \autoref{fig:reach_results_swe} and \autoref{fig:reach_results_sales})
  the gap between gender delivery for each pair of ads  
   is reduced for the \emph{reach} ads compared to the \emph{conversion} ads.
And, for two of the job categories (delivery driver and sales associate),
  one of the three cases does not show a statistically significant evidence for skew,
  while all three showed such evidence in the \emph{conversion} ads.
These observations indicate that the degree of skew may be reduced when using the \emph{reach} objective,
  and, therefore, an advertiser's request for the \emph{conversion} objective
  may increase the amount of skew because, according to Facebook's algorithmic predictions, conversions may correlate with
  particular gender choices for certain jobs.

Revisiting our discussion of the legal responsibility for discrimination (\autoref{sec:liability})
  in light of these results,
  the fact that both the advertiser and the platform make choices
  about the ad recipients can blur who is legally responsible.
If the discriminatory outcome occurs regardless of the advertiser-chosen objective,
  as our results with the \emph{reach} objective underscore,
  then it's clear it is the responsibility of the platform.
On the other hand, if we saw skew with advertiser-specified objectives
  that optimize for engagement and not others (which was not the case in our experiments),
  the platform may claim it is just doing what the advertiser requested,
  and may even state that the blame (or legal culpability) for any skew
  therefore rests on the advertiser.  
However, even in this case, one could argue that it is the ad platform that has full 
control over determining how the optimization algorithm actually works and what its inputs are.
Therefore, 
if an advertiser discloses that they are running job ads (which we did in our experiments),
  the ad platform may still have the ethical and legal responsibility to ensure
  its algorithm does not produce a discriminatory outcome
  regardless of the advertiser objective it is optimizing for.

\section{Future Work}
  \label{sec:discussion}

We next discuss the limitations of our study,
  give some directions for future study and, motivated by the challenges
  we faced in our work, provide recommendations as to what ad platforms
  can do to make auditing more feasible and accurate.

\subsection{Limitations and Further Directions}
	\label{sec:study_directions}

Our experiments focus on skew from gender,
  but we believe our methodology can be used to study other attributes such as age or race.
It requires the auditor having access to data about age and gender distributions among employees of different companies in the same category, 
so that the auditor can pick job ads that fit the criteria of our methodology.
It also requires the ability to create audiences whose age and race distributions are known.
The voter dataset we use includes age and race, so can
  be adapted to test for discrimination along those attributes.

Like prior studies, we use physical location as a proxy to infer the gender
  of the ad recipient, an approach which has some limitations.
LinkedIn hides location when there are two or fewer ad recipients,
  so our estimates may be off in those areas.
These cases account for 31-43\% of our ad recipients,
  as shown in \autoref{tab:gender_breakdown}.  
Assuming gender distribution is uniform by county in North Carolina's population,
  we reason that these unreported cases do not significantly distort our conclusions.

We tested three job categories, with three experiment repetitions each.
Additional categories and repetitions would improve confidence in our results.
Although we found it difficult to select job categories
  with documented gender bias that we could target,
  such data is available in private datasets.
Another question worth investigating with regards to picking job categories is whether
  delivery optimization algorithms are the same for all job categories, i.e., whether
  relatively more optimization happens for high-paying or scarce jobs.

Some advertisers will wish to target their ads by profession or background.
We did not evaluate such targeting because our population data
  is not rich and large enough to support such comparisons with
  statistical rigor.
Evaluation of this question would be future work, especially if the auditor has access to richer population data.

\begin{table}
\caption{Breakdown of impressions for LinkedIn ads run using Aud\#2. ``Unreported'' shows
 percentages in unreported counties (whose genders we cannot infer).}
\label{tab:gender_breakdown}
\centering
\begin{tabular}{ |c|c|c|c|c|c| }
\hline
Company & Total & Males & Females & Unreported (\%) \\ 
\hline
Domino's & 806 & 241 & 233 & 41.19 \\
Instacart & 757 & 194 & 232 & 43.73 \\
Nvidia & 859 & 232 & 258 & 42.96 \\
Netflix & 907 & 240 & 272 & 43.55 \\
Leith & 454 & 145 & 160 & 32.82 \\
Reeds & 521 & 192 & 166 & 31.29 \\
\hline
\end{tabular}
\end{table}

\subsection{Recommendations}
  \label{sec:recommendations}

Prior work has shown that platforms are not consistent when self-policing
  their algorithms for undesired societal consequences,
  perhaps because the platforms' business objectives are at stake.
Therefore, we believe independent (third party)
  auditing fills an important role.
We suggest recommendations
  to make
    such external auditing of ad delivery algorithms more accessible, accurate and efficient, especially for public interest researchers and journalists.

\textbf{Providing more targeting and delivery statistics:}
First, echoing sentiments from prior academic and activism work~\cite{Ali2019a, Mozilla}, we note the value of surfacing additional ad targeting and delivery data in a privacy-preserving way.
Public interest auditors often rely on features that the ad platforms make available
  for any regular advertiser to conduct their studies, which can make performing
  certain types of audits challenging.
For example, in the case of LinkedIn, the ad performance report does not contain
  a breakdown of ad impressions by gender or age.
To overcome such challenges, prior audit studies and our work rely on finding workarounds such as
  proxies to measure ad delivery along sensitive demographic features.
On one hand, providing additional ad delivery statistics could help expend the scope of auditors' investigations.
On the other hand, there may be an inherent trade-off between providing additional statistics about
  ad targeting and delivery and the privacy of users (see e.g.~\cite{Korolova2010, Faizullabhoy2018}) or business interests of advertisers.
We believe that privacy-preserving techniques, such as differentially private data publishing~\cite{dwork2014algorithmic} may be able to strike a balance between auditability and privacy, and could be a fruitful direction for future work and practical implementation in the ad delivery context.
It is also worth asking what additional functionalities or insights about its data or ad delivery optimization algorithms
  the platforms can or should provide which
  would allow for more accessible auditing without sacrificing independence of the audits.
Recent work has explored finding a balance between independence and addressing
  the challenges of external auditing by suggesting a \emph{cooperative audit} framework~\cite{wilson2021building},
  where the target platform is aware of the audit and gives the auditor special access
  but there are certain protocols in place to ensure the auditor's independence.
In the context of ad platforms, we recognize that providing a special access option for auditors
  may open a path for abuse where advertisers may pretend to be an auditor for their economic or competitive benefit.

\textbf{Replacing ad-hoc privacy techniques:}
Our other recommendation is for ad platforms to replace ad-hoc techniques they
  use as a privacy enhancement with more rigorous approaches.
For example, LinkedIn gives only a rough estimate of audience sizes,
  and does not give the sizes if less than 300~\cite{AudienceCounts}
It also does not give the number of impressions by location if the count per county is less than three~\cite{AdsReporting}.

Such ad-hoc approaches have two main problems. First, it is not clear
   based on prior work on the ad platforms how effective they are in terms of protecting privacy of users~\cite{Venkatadri2018, venkatadri-2020-composition}.
We were also able to circumvent the 300-minimum limit for audience size estimates on LinkedIn
  with repeated queries by composing one targeting parameter with another,
  then repeating a decomposed query and calculating the difference.
More generally, numerous studies show ad-hoc approaches often
  fail to provide the privacy that they promise~\cite{Netflix, cohen2018linear}.
Second, ad-hoc approaches can distorts statistical tests that
  auditors perform~\cite{nissim2017differential}.
Therefore, we recommend ad platforms use approaches with rigorous
  privacy guarantees, and whose impact on statistical validity can be precisely analyzed, such as differentially private algorithms~\cite{dwork2014algorithmic}, where possible.

\textbf{Reducing cost of auditing:}
Auditing ad platforms via black-box techniques incurs a substantial cost of
  money, effort, and time.
Our work alone required several months of research on data collection and methodology
  design, and cost close to \$5K to perform the experiments by running ads.
A prior study of the impact of Facebook's ad delivery algorithms on political discourse cost up to \$13K~\cite{Ali2019b}.
These costs quickly accumulate if one is to repeat experiments
  to study trends, increase statistical confidence, or reproduce results.
One possible solution is to provide a discount for auditors.
They would have similar access to the platform features like any other
  advertiser but would pay less to run ads.
However, as with other designed-auditor techniques,
  this approach risks abuse.

Overall, making auditing ad delivery systems more feasible to a broader range of interested parties can help ensure that the systems that shape job opportunities
  people see operate in a fair manner that does not violate anti-discrimination laws.
The platforms may not currently have the incentives to make the changes proposed and, in some cases, may actively block transparency efforts initiated by researchers and journalists~\cite{Merrill2019}; thus, they may need to be mandated by law.

\section{Conclusion}
We study gender bias in the delivery of job ads due to platform's optimization choices,
  extending existing methodology to account for the role of
qualifications in addition to the other confounding factors studied in prior work.
We are the first to methodologically address the challenge of controlling
  for qualification, and also draw attention to how qualification may be used as a
  legal defense against liability under applicable laws.
We apply our methodology to both Facebook and LinkedIn
  and show that our proposed methodology is applicable to multiple platforms and 
can identify distinctions between their ad delivery practices.
We also 
  provide the first analysis of LinkedIn for potential skew in ad delivery.
\basi{}
We confirm that Facebook's ad delivery can result in skew of job ad delivery by gender
  beyond what can be legally justified by possible differences in qualifications, thus strengthening the previously raised arguments that Facebook's ad delivery algorithms may be in violation of anti-discrimination laws~\cite{datta2018discrimination, Ali2019a}.
We do not find such skew on LinkedIn.
Our approach provides a novel example of feasibility of auditing algorithmic systems in a black-box manner, using only the capabilities available to all users of the system.
At the same time, the challenges we encounter lead us to suggest changes that ad platforms could make (or that should be mandated of them)
  to make external auditing of their performance in societally impactful areas easier.

\section*{Acknowledgements}
This work was funded in part by NSF grants CNS-1755992, CNS-1916153, CNS-1943584, CNS-1956435,
  and CNS-1925737. %
We are grateful to Aaron Rieke for feedback on the manuscript.
\balance
\bibliographystyle{acm}

%
%
%
%
%
%
%
%
%
%
%
%
%
%
%
%
%
%
%
%
%
%
%
%
%
%
%
%
%
%
%
%
%
%
%
%
%
%
%
%
%
%
%
%
%
%
%
%
%
%
%
%
%
%
%
%
%
%
%
%
%
%
%
%
%
%
%
%
%
%
%
%
%
%
%
%
%
%
%
%
%
%
%
%
%
%
%
%
%
%
%
%
%
%
%
%
%
%
%

\end{document}